\documentclass[reqno]{amsart}

\usepackage{graphicx}
\usepackage{sidecap}
\usepackage{color}
\usepackage{natbib}
\usepackage{hyperref}
\hypersetup{colorlinks,urlcolor=blue,citecolor=blue}

\usepackage{listings}

\theoremstyle{definition}

\theoremstyle{remark}

\numberwithin{equation}{section}

\DeclareMathOperator{\COV}{Cov}

\DeclareMathOperator{\diag}{diag}
\DeclareMathOperator{\E}{E}

\def\Rset{\mathbb{R}}
\DeclareMathOperator{\tr}{tr}
\DeclareMathOperator{\VAR}{Var}

\def\Zset{\mathbb{Z}}

\newcommand{\bm}[1]{\mbox{\boldmath $#1$}} 
\newcommand{\what}[1]{\widehat{#1}}

\DeclareGraphicsExtensions{.pdf,.jpg}

\begin{document}

\title[SpatialPack: Computing the Spatial Association]{\textsf{SpatialPack}:
  Computing the Association Between Two Spatial Processes}

\author{Felipe Osorio}
\address{Instituto de Estad\'istica, Pontificia Universidad Cat\'olica de Valpara\'iso, Chile}
\curraddr{Avenida Err\'azuriz 2734, Valpara\'iso, Chile}
\urladdr{\includegraphics[scale=.12]{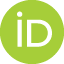}\,\url{http://orcid.org/0000-0002-4675-5201}}
\email{\href{mailto:felipe.osorio@pucv.cl}{felipe.osorio@pucv.cl}}

\author{Ronny Vallejos}
\address{Departamento de Matem\'atica, Universidad T\'ecnica Federico Santa Mar\'ia}
\curraddr{Avenida Espa\~na 1680, Valpara\'iso, Chile}
\email{\href{mailto:ronny.vallejos@usm.cl}{ronny.vallejos@usm.cl}}

\author{Francisco Cuevas}
\address{Department of Mathematical Sciences, Aalborg University}
\curraddr{Fredrik Bajers Vej 7, Aalborg, Denmark}
\email{\href{mailto:francisco@math.aau.dk}{francisco@math.aau.dk}}

\maketitle

\begin{abstract}
An R package \textsf{SpatialPack} that implements routines to compute point estimators
and perform hypothesis testing of the spatial association between two stochastic
sequences is introduced. These methods address the spatial association between two
processes that have been observed over the same spatial locations. We briefly review
the methodologies for which the routines are developed. The core routines have been
implemented in \texttt{C} and linked to R to ensure a reasonable computational speed.
Three examples are presented to illustrate the use of the package with both simulated
and real data. The particular case of computing the association between two time
series is also considered. Besides elementary plots and outputs we also provide a
plot to visualize the spatial correlation in all directions using a new graphical
tool called codispersion map. The potential extensions of \textsf{SpatialPack} are
also discussed.
\end{abstract}

\section{Introduction}

The increasing need to analyze the large dimension data in spatial modeling highlights
the necessity of having suitable and efficient routines to perform the spatial data
analysis. For this and other reasons, researchers have been motivated to create many
R packages that contain routines and functions that facilitate the computation of the
standard procedures and the new theory developed in spatial statistics. Examples of
commonly used packages in spatial analysis and geostatistics are \textsf{geoR} \citep{Ribeiro:2001},
\textsf{spatstat} \citep{Baddeley:2005}, and \textsf{GeoXp} \citep{Laurent:2012}, among
others. A complete set of R functions developed in the context of spatial econometrics
can be found in \cite{Bivand:2002}.

In the analysis of spatial data, the quantification of spatial associations between
two variables is an important issue, and considerable effort has been devoted to the
construction of appropriate coefficients and tests for the association between two
correlated variables. One can approach this issue by removing the spatial association
among the observations and then applying existing techniques that have been developed
for independent variables. An alternative way to manage the problem is to consider
approaches that allow one to take into account the autocorrelation structure of the data.

Here, we describe a new R package \textsf{SpatialPack} that adds techniques to assess
the spatial association between the two processes defined on a finite subset of the
plane/real line. The techniques implemented in the package consist of two coefficients
of association introduced by Tj{\o}stheim and Matheron, respectively \citep{Tjostheim:1978,
Matheron:1965} and one hypothesis testing procedure studied by \cite{Clifford:1989}.
These three techniques tackle the spatial association between the two sequences defined
in the same locations on the plane and are part of the standard procedures used to
analyze the relationship between two spatial variables. Applications of these methodologies
in several different disciplines can be found in \cite{Goovaerts:1997, Pringle:2006,
Blanco-Moreno:2005, Vallejos:2012} and \cite{Ojeda:2012}. Although these methods are
part of the existing techniques to assess the spatial association between two spatial
variables, they are still considered in current research \citep{Dutilleul:2008, Cuevas:2013}.

Tj{\o}stheim's coefficient is a nonparametric coefficient constructed from the ranks
of suitable transformations of the coordinates for which both processes are defined.
Matheron's coefficient is also known as the codispersion coefficient and is a measurement
of association that depends on a distance lag $\bm{h}$. The codispersion coefficient
is a normalization of the widely used cross-variogram and shares several properties
with the correlation coefficient; however, the difference between the two coefficients
relies on the fact that the codispersion coefficient is not centered by the process
means. Instead, the coefficient quantifies the expected value of the cross-product
between observations that are separated by a distance $\bm{h}$. In this sense, the
correlation coefficient is a crude measurement of spatial association because it does
not depend on a given direction $\bm{h}$ on the plane. The hypothesis testing described
in \cite{Clifford:1989} is based on a modified version of the correlation coefficient,
which can be applied to regular or irregular grids.

Three functions were developed to compute the procedures described above: \texttt{cor.spatial},
\texttt{codisp}, and \texttt{modified.ttest}. All of the basic routines were built in
\texttt{C} for efficiency and then properly linked to R. Examples with real and simulated
datasets are presented to illustrate the capabilities of the techniques implemented in
the package. Under specific correlation structures between and across two spatial processes,
a Monte Carlo simulation study was conducted to explore the computational time as a
function of the sample size. Two examples with real data are discussed to illustrate
practical applications. The first example demonstrates the well-known Murray smelter
site dataset in which the variables of interest are arsenic (As) and lead (Pb), both
defined on a non-regular grid on the plane. The second example relates to the flammability
of polymers previously studied by \cite{Rukhin:2008}. Our analysis provides a more
complete discussion about the similarity of four images used in the previous study.
A brief discussion about comovement in time series, including a real data example is
presented. The construction of a new graphical tool called codispersion map \citep{Vallejos:2015b}
is outlined in the discussion. This map allows us to visualize the correlation between
two spatial processes in all directions of interest in a single graph. Concluding
remarks and possible topics for further research are also provided.

The package is available from the Comprehensive R Archive Network at \url{http://CRAN.R-project.org/package=SpatialPack}.
New versions of the package, which is still under development, will be available at
\url{http://www.ies.ucv.cl/spatialpack/}.

\section{Methods}

The methodology discussed in this section includes three different approaches to address the spatial association
between two stochastic sequences indexed on a $d$-dimensional space. In all cases we consider $D \subset\Rset^d$,
$\{X(\bm{s}):\bm{s}\in D\}$ and $\{Y(\bm{s}): \bm{s} \in D\}$ are two spatial processes and the available data
at the spatial locations $\bm{s}_1,\bm{s}_2,\dots,\bm{s}_n \in D$ are the pairs $(X(\bm{s}_i), Y(\bm{s}_i))$, $i=1,2,
\dots,n$. Throughout the paper we assume that $d=2$ with the exception of Example 4 in which $d=1$.

\subsection{The modified correlation coefficient}

\cite{Clifford:1989} developed tests of association between two spatially correlated processes. These
tests are based on modifying the variance and degrees of freedom of the standard $t$-test and required the estimation
of the effective sample size. The later one is the factor that takes into account the spatial association of both processes.
Here, we briefly described their method.

Let us consider $A\subset D$ a set of $n$ locations, say $A=\{\bm{s}_1,\bm{s}_2,\dots,\bm{s}_n\}$. Suppose that $\bm{X}=
(X(\bm{s}_1),X(\bm{s}_2),\dots,X(\bm{s}_n))^\top$ and $\bm{Y}=(Y(\bm{s}_1),Y(\bm{s}_2),\dots,Y(\bm{s}_n))^\top$ are multivariate
normal vectors with constant means and covariance matrices $\bm{\Sigma}_X$ and $\bm{\Sigma}_Y$, respectively.

Assume that $D$ can be divided into strata $D_0,D_1,D_2,\dots$, so that $\COV(X(\bm{s}_i),X(\bm{s}_j))=C_X(k)$ and
$\COV(Y(\bm{s}_i),Y(\bm{s}_j)) = C_Y(k)$, with $\bm{s}_i,\bm{s}_j \in D_k$, for $k=0,1,\dots$. \cite{Clifford:1985}
have suggested using
\[
 \widehat{C}_Y(h) = \sum_{\bm{s}_i, \bm{s}_j \in A_k} (Y(\bm{s}_i)-\overline{Y})(Y(\bm{s}_j)-\overline{Y})/n_k,
\]
as an estimate of $C_Y(h)$, where $n_k$ is the the cardinality of $D_k$ and similarly for $C_X(k)$. Later
\cite{Clifford:1989} suggested to use
\begin{equation}\label{var1}
 n^{-2}\sum_{h}n_h \widehat{C}_X(h) \widehat{C}_Y(h)
\end{equation}
as an estimate of the conditional variance of $s_{XY}=n^{-1}\sum_D (X(\bm{s})-\overline{X})(Y(\bm{s})-\overline{Y})$. As a result
the  modified $t$-test proposed in \cite{Clifford:1989} is based on the statistic
\begin{equation}\label{eq:W}
 W = n\, s_{XY}\left(\sum_{h}n_h \widehat{C}_X(h) \widehat{C}_Y(h)\right)^{-2}.
\end{equation}
Performing some approximations to the variance of the correlation coefficient $\sigma_r^2$ between processes $X(\bm{s})$
and $Y(\bm{s})$ see Appendix 1 in \cite{Clifford:1989} the test statistic $W$ can be written as
\begin{equation}\label{eq:W2}
 W = (\widehat{M}-1)^{1/2}r,
\end{equation}
where $r$ is the correlation coefficient between $X(\bm{s})$ and $Y(\bm{s})$, $\widehat{M}=1+\widehat{\sigma}^{-2}_r$,
and
\[
 \widehat{\sigma}^2_r = \frac{\sum_{h}n_h \widehat{C}_X(h) \widehat{C}_Y(h)}{n^2s_X^2s_Y^2}.
\]
The test statistic (\ref{eq:W2}) was studied assuming that under the null hypothesis of no spatial correlation between processes
$X(\bm{s})$ and $Y(\bm{s})$, $W$ has a $t$-student distribution with $\widehat{M}-2$ degrees of freedom. Further discussions
and extensions of (\ref{eq:W2}) can be found in \cite{Dutilleul:1993} and \cite{Dutilleul:2008}.

\subsection{Tj{\o}stheim's coefficient}

\cite{Tjostheim:1978} introduced a nonparametric coefficient to measure the association between two spatial sequences.
Following the notation introduced above, consider again two spatial processes $X(\bm{s})$ and $Y(\bm{s})$, defined on a two-dimensional
space, that is $\bm{s}=(s_1,s_2)$. Define the function
\begin{equation}\label{c}
 c(u) = \begin{cases}
 0, & u<0, \\
 1, & u>0, \\
 \frac{1}{2}, & u=0.
 \end{cases}
\end{equation}
Then the rank $R(\bm{s}_i)$ of process $X(\bm{s})$ at the point $\bm{s}_i$ is defined as
\begin{equation}\label{eq:rank}
 R_X(\bm{s}_i)=\sum_{j=1}^n c(X(\bm{s}_i)-X(\bm{s}_j)), \qquad i\neq j,
\end{equation}
and similarly for $R_Y(\bm{s}_i).$ We define the $s_1$ coordinate corresponding to the rank $i$ of $X(\bm{s})$ as $F_X$. Then
\begin{equation}\label{eq:coor-rank}
 F_X(i) = \sum_{j=1}^n s_{j1}\delta(i,R_X(\bm{s}_i)),
\end{equation}
where $\delta(i,j)$ is the Kronecker delta and $\bm{s}_i=(s_{i1},s_{i2})$, $i=1,\dots,n$. The quantities $G_X(i)$, $F_Y(i)$ and $G_Y(i)$ are
defined similarly.

Tj{\o}stheim's coefficient is defined as
\begin{equation}\label{eq:A}
 A = \frac{\sum_i [(F_X(i)-\overline{F})(F_Y(i)-\overline{F})][(G_X(i)-\overline{G})
 (G_Y(i)-\overline{G})]}{\sqrt{\sum_i [(F_X(i)-\overline{F})^2+(G_X(i)-\overline{G})^2]
 \sum_i [(F_Y(i)-\overline{F})^2+(G_Y(i)-\overline{G})^2]}},
\end{equation}
where $\overline{F}=\sum_i F_X(i)/n$ and similarly for $\overline{G}$. \cite{Tjostheim:1978} proved that if $\bm{X}$
and $\bm{Y}$ are two vectors of $n$ independent random variables and $\bm{X}, \bm{Y}$ are also independent, then the variance of
$A$ is given by
\begin{equation}\label{eq:varA}
 \VAR(A)=\frac{(\sum_i s_{i1}^2)^2+2(\sum_i s_{i1}s_{i2})^2+(\sum_i s_{i2}^2)^2}{(n-1)(\sum_i s_{i1}^2 +\sum_i s_{i2}^2)^2}.
\end{equation}
A discussion about the advantages and drawbacks of Tj{\o}stheim's coefficient and some extensions are in \cite{Hubert:1982}.

\subsection{The codispersion coefficient}

Let $X(\bm{s})$ and $Y(\bm{s})$ two intrinsically stationary processes. The cross variogram between $X(\bm{s})$ and $Y(\bm{s})$ is defined as
\begin{equation}\label{crossvar}
 2\gamma_{XY}(\bm{h}) = \E[(X(\bm{s}+ \bm{h}) - X(\bm{s}))(Y(\bm{s}+\bm{h}) - Y(\bm{s}))],
\end{equation}
for all $\bm{s},\bm{s}+\bm{h} \in D$. The codispersion coefficient studied by \cite{Matheron:1965} is a normalized version of (\ref{crossvar}) given by
\begin{equation}\label{eq:rho}
 \rho_{XY}(\bm{h}) = \frac{\E[(X(\bm{s}+\bm{h}) - X(\bm{s}))(Y(\bm{s}+\bm{h}) - Y(\bm{s}))]}
 {\sqrt{\E[(X(\bm{s}+\bm{h}) - X(\bm{s}))^2]\E[(Y(\bm{s}+\bm{h}) - Y(\bm{s}))^2]}}
 = \frac{\gamma_{XY}(\bm{h})}{\sqrt{\gamma_X(\bm{h})\gamma_Y(\bm{h})}}.
\end{equation}
A method of moment estimator of the codispersion coefficient is
\begin{equation}\label{eq:codisp}
 \widehat{\rho}_{XY}(\bm{h}) = \frac{\sum_{N(\bm{h})}(X(\bm{s}_i)-X(\bm{s}_j))(Y(\bm{s}_i)-Y(\bm{s}_j))}
 {[\sum_{N(\bm{h})}(X(\bm{s}_i)-X(\bm{s}_j))^2 \sum_{N(\bm{h})}(Y(\bm{s}_i)-Y(\bm{s}_j))^2]^{1/2}},
\end{equation}
where $N(\bm{h})=\{(\bm{s}_i,\bm{s}_j): \bm{s}_i-\bm{s}_j=\bm{h}, 1\leq i,j\leq n\}$. Under very precise conditions the consistency
and asymptotic normality of (\ref{eq:codisp}) for spatial autoregressive processes were studied by Rukhin and Vallejos \cite{Rukhin:2008}.
Vallejos \cite{Vallejos:2008} adapted the results for time series models and \cite{Ojeda:2012} used the codispersion coefficient
as a measure of similarity between images.

In practice the codispersion coefficient is plotted in a similar way as the correlation function in time series. Creating bins as in
the variogram computation for isotropic processes \citep{Banerjee:2004}, define
\begin{equation}\label{eq:codisp2}
 \widehat{\rho}_{XY}(h_k) = \frac{\sum_{(\bm{s}_i, \bm{s}_j)\in N(h_k)}(X(\bm{s}_i)-X(\bm{s}_j))(Y(\bm{s}_i)-Y(\bm{s}_j))}
 {[\sum_{(\bm{s}_i,\bm{s}_j)\in N(h_k)}(X(\bm{s}_i)-X(\bm{s}_j))^2 \sum_{(\bm{s}_i,\bm{s}_j)\in N(h_k)}(Y(\bm{s}_i)-Y(\bm{s}_j))^2]^{1/2}},
\end{equation}
where $N(h_k)=\{(\bm{s}_i,\bm{s}_j): \|\bm{s}_i-\bm{s}_j\| \in I_k\}$ for $k=1,2,\ldots, K,$ $I_k$ indexes the $k$th bin.

\section{Software implementation in R}

The code that supports the R package \textsf{SpatialPack} is described in this section. The set of routines correspond to the
methods of quantifying the spatial association between the two processes introduced in the previous Section. In the current version
of the package, three functions to compute the coefficients and tests described before are available: \texttt{modified.ttest},
\texttt{cor.spatial} and \texttt{codisp}. To achieve computational efficiency, each function calls a dynamic link library (shared
library) written in \texttt{C}. Internally, some basic linear algebra subprograms (BLAS) \citep{Lawson:1979} that are included in
R have been used to perform matrix operations. The routines that compute the modified $t$-test and the codispersion coefficient
share a number of structures that allow the storage of intermediate results. For these routines, there is an initialization step
in which the distances between the locations and the maximum distance are computed. Subsequently, the upper bounds for each bin
are computed. To preserve the minimum storage requirements, we avoided storing the matrix of distances. Instead, the distances were
computed directly (when they were required). This strategy facilitates the treatment of large datasets. The arguments of the
implemented functions are an $n\times 2$ matrix containing the coordinates where the observations were measured and two $n$-dimensional
vectors \texttt{x} and \texttt{y} that contain the observations for the first and second spatial variable, respectively. In addition,
the number of classes (bins) can be specified through the argument \texttt{nclass}, otherwise (default) 13 classes are constructed
similarly to the \texttt{variog} function in the \textsf{geoR} package \citep{Ribeiro:2001}. Alternatively, Sturges' formula
\citep{Sturges:1926} can be used to implicitly base the bin sizes on the data range.

Next, the base functions of \textsf{SpatialPack} are described.

\subsection{Function \textsf{modified.ttest}}
The modified $t$-test \citep{Clifford:1989}, which measures the spatial association between two variables, has been computationally
implemented in the function \texttt{modified.ttest}. To test the null hypothesis of no spatial correlation between $X(\bm{s})$ and
$Y(\bm{s})$, the $F=(\what{M}-2)r^2/(1-r^2)$ test function has been used, which (under the null hypothesis) follows an $F$ distribution
with 1 and $\what{M}-2$ degrees of freedom. The \texttt{C} code underlying the function \texttt{modified.ttest} computes an effective
sample size $\what{M}=1+\what{\sigma}_r^{-2}$ where $\what{\sigma}_r^{-2}$ is computed from the expression for $\sigma_r^2$ provided
by \cite{Dutilleul:1993}:
\[
 \what{\sigma}_r^2 = \frac{\tr(\bm{B}\what{\bm{\Sigma}}_X\bm{B}\what{\bm{\Sigma}}_Y)}
 {\tr(\bm{B}\what{\bm{\Sigma}}_X)\tr(\bm{B}\what{\bm{\Sigma}}_X)},
\]
where $\bm{B}=\bm{I}_n-\frac{1}{n}\bm{1}_n\bm{1}_n^\top$ is the centering matrix of order $n$, $\bm{I}_n$ is the identity matrix of order
$n$, and $\bm{1}_n$ is an $n$-dimensional vector of ones. The covariance matrices $\bm{\Sigma}_X$ and $\bm{\Sigma}_Y$ are estimated using
Moran's index \citep{Moran:1950}. The routine does not build the matrix $\bm{B}$; however, it takes advantage of its structure to simplify
certain calculations. The output of the function \texttt{modified.ttest} is an object of the \texttt{mod.ttest} class, which has components
that include the statistic $F$ (\texttt{Fstat}), the estimated degrees of freedom (\texttt{dof}), the $p$-value associated to the test
(\texttt{p.value}), the upper bounds for the classes (\texttt{upper.bounds}), the number of observations in each class (cardinality)
(\texttt{card}), and an $K\times 2$ matrix containing the Moran indices (\texttt{imoran}) for each variable under study (where $K$ is the
number of classes). All of this information is appropriately displayed in the output when using the methods \texttt{print} and \texttt{summary}.

Matlab code related to the computation of  the modified $t$-test (following Dutilleul's guidelines)  can be found on the website
\url{http://environmetricslab.mcgill.ca/Programs.html}.

\subsection{Function \textsf{cor.spatial}}

The coefficient first introduced by \cite{Tjostheim:1978}  was implemented through the \texttt{cor.spatial} function. This procedure
handles the possible ties that can occur in the observed values through the option \texttt{ties.method = "first"}. This method is an
existing option of function \texttt{rank}, which is available in R. In the computation of Tj{\o}stheim's coefficient, the coordinates
of the ranks defined in equation (\ref{eq:coor-rank}) are first centered and then computed using R commands, while the computation of
$A$ in equation (\ref{eq:A}) is performed in \texttt{C}. Internally, the calculations are optimized by calling level 1 routines from
BLAS \citep{Lawson:1979}. The procedure also returns $\VAR(A)$ as an attribute of \texttt{"variance"}.

\subsection{Function \textsf{codisp}}

The \texttt{codisp} function computes the codispersion coefficient for the general (non-rectangular) grids according to its definition
in the equation (\ref{eq:codisp}). This function shares some \texttt{C} code developed for the function \texttt{modified.ttest}. The
output object corresponds to a class list \texttt{"codisp"} whose components have a structure similar to the elements that were defined
for the modified $t$-test. The value of the output \texttt{coef} corresponds to a vector of size \texttt{nclass} that contains the values
of the codispersion coefficient $\what{\rho}_{XY}(h_k)$ for each one of the strata previously defined. The information associated with
the upper bounds of the strata is returned in \texttt{card} from the output subject. A generic function has been written to appropriately
print the results that are obtained by the function \texttt{codisp}. Additionally, $\what{\rho}_{XY}(h_k)$ versus $h_k$ can be plotted
using the \texttt{plot} method.

\section{\textsf{SpatialPack} in practice}

In this section three examples are introduced. The first one uses simulated data to inspect the capabilities of the methods described above. The second
one works with a real dataset defined on a non-rectangular grid. The third example describes a dataset that consist of four images defined on a regular grid.

\subsection{Example 1: A Monte Carlo simulation study}

To explore the computational time associated with the functions that compute the codispersion coefficient and the modified $t$-test, a Monte Carlo simulation
experiment was conducted. Enlarged images sized $8\times 8,$ $16\times 16,$ $32\times 32,$ and $64 \times 64$ were generated to study the behavior of the
procedures \texttt{modified.ttest} and \texttt{codisp} as a function of the image size.

\begin{figure}[!htp]
\begin{center}
\begin{minipage}[c]{6cm}
\includegraphics[scale=.33]{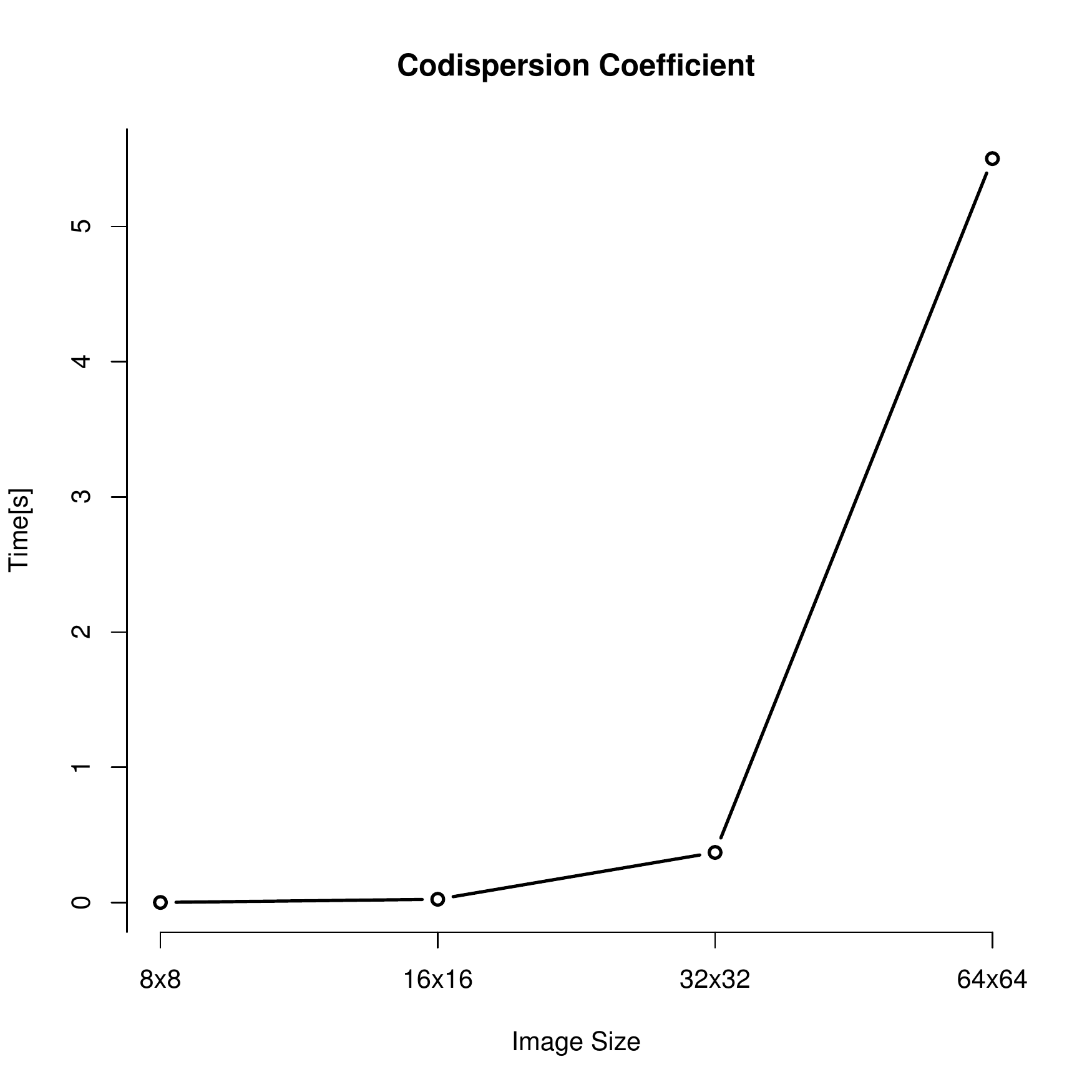}
\end{minipage}
\begin{minipage}[c]{6cm}
\includegraphics[scale=.33]{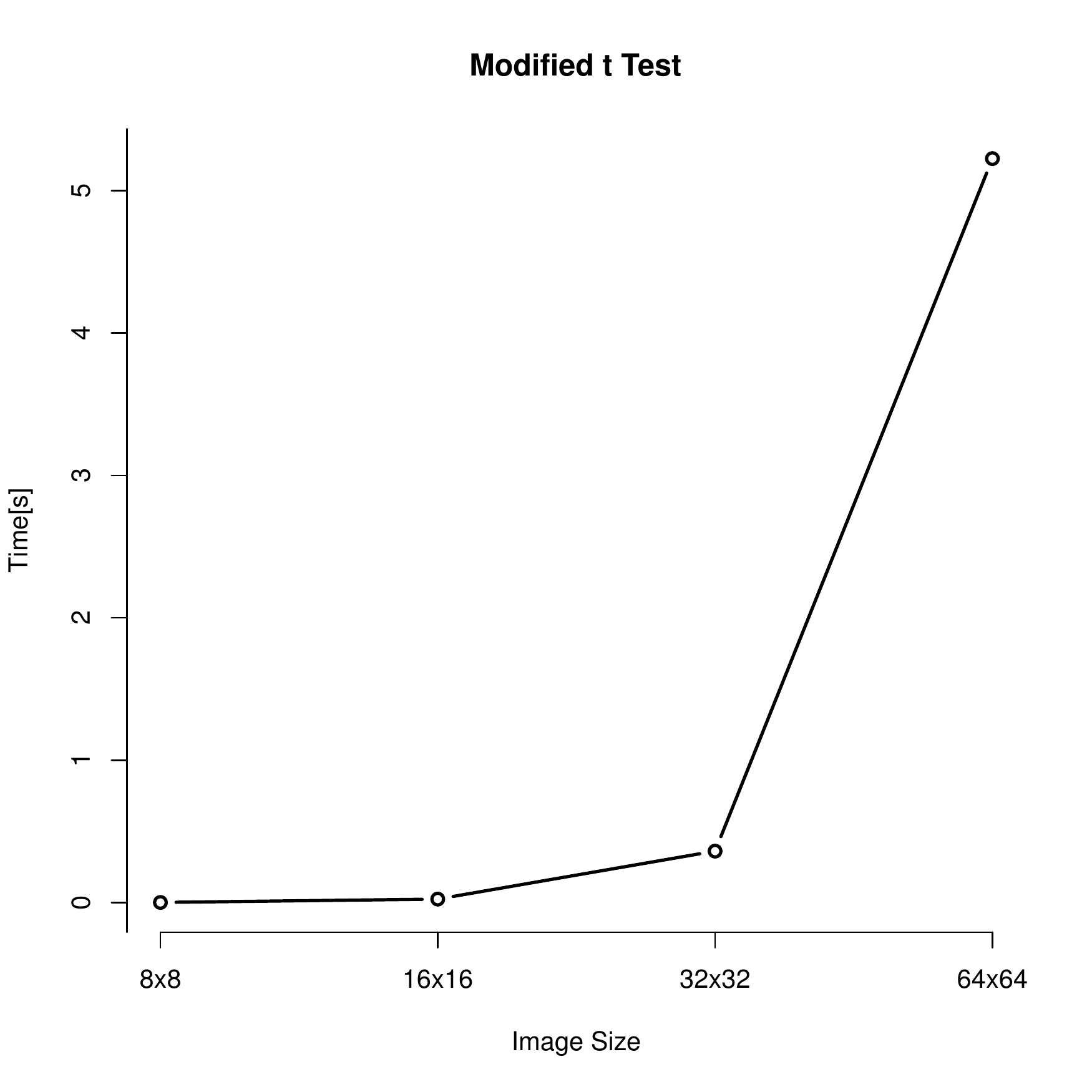}
\end{minipage}
\end{center}
\caption{computational time to evaluate the codispersion coefficient (left) and the modified \emph{t} test (right) as a function of the image size
(from $8\times 8$ to $64 \times 64$).}
\label{time}
\end{figure}

The generated images were the realizations of two correlated spatial processes. Precisely, let $X(\bm{s})$ and $Y(\bm{s})$ be two spatial processes defined
for $\bm{s}\in D\subset \Zset^d$. For the locations $\bm{s}_1,\bm{s}_2,\dots,\bm{s}_n$, consider the processes $\bm{X}=(X(\bm{s}_1), X(\bm{s}_2),
\dots,X(\bm{s}_n))^\top$ and $\bm{Y}=(Y(\bm{s}_1),Y(\bm{s}_2),\dots,Y(\bm{s}_n))^\top$ and define the process $\bm{Z}=(\bm{X}^\top,\bm{Y}^\top)^\top
\sim\mathcal{N}_{2n}(\bm{0},\bm{\Sigma}),$ where the elements of the covariance matrix $\bm{\Sigma}$ are given by:
\[
 \bm{\Sigma}_{ij} = \begin{cases}
 C_0(\|\bm{s}_i-\bm{s}_j\|), & 1\leq i,j\leq n, \\
 C_1(\|\bm{s}_i-\bm{s}_j\|), & 1\leq i\leq n, \ n+1\leq j \leq 2n, \\
 C_0(\|\bm{s}_i-\bm{s}_j\|), &  n+1\leq i, j \leq 2n,
 \end{cases}
\]
where $C_u(\bm{h})=C(\bm{h},u),$ with $C(\bm{h},u)$ a nonseparable covariance function on the domain $\Rset^d \times \Rset$, which belongs to the
parametric family \citep[see][]{Gelfand:2011}
\[
 C(\bm{h},u) = \frac{1}{\psi(u^2)^{d/2}}\,\eta\left(\frac{\|\bm{h}\|^2}{\psi(u^2)}\right),
\]
where $\psi(r)=(a r^\alpha+1)^\beta$, $\eta(r)=(1+(r/\sigma^2)^\gamma)^{-c/\gamma}$, $\alpha\in(0,1]$, $\beta\in(0,1]$, $a>0$, $c>0$, and
$0<\gamma\leq 2$. In the simulation study, four images sized $8\times 8, 16\times 16, 32\times 32,$ and $64\times 64$ were considered. The corresponding
sample sizes associated with these images are $n=8^2,16^2,32^2,64^2$. The following set of parameters was chosen: $a=\alpha=\beta=\sigma=1$, $c=3,
\gamma=2.$ The process $\bm{Z}$ was generated 10 times for each image and the computational time of procedures \texttt{modified.ttest} and \texttt{codisp}
was recorded. The average times for each function are shown in Figure \ref{time}.

We recall that each procedure requires the computation of $n(n-1)/2$ distances for each bin. Considering $\texttt{nclass}=13$, the total number of operations
required to compute either \texttt{codisp} or \texttt{modified.ttest} for the two images sized $64 \times 64$ is $(\texttt{nclass}+1) n(n-1)/2=117\, 411\, 840$.
For the function \texttt{modified.ttest}, $n^2=16\, 777\, 216$ extra operations are needed to evaluate the test statistic. Although a number of operations are
required to compute these functions, in practice the computational time remains reasonable. For example, for an image sized $64\times 64$ both procedures take
approximately 5 seconds in a PC with a Core 2 quad q8400 2.66 Ghz processor, and 8 Gb RAM DDR2 800 MHz (see Figure \ref{time}). This feature is due to the
strategy used to organize the computations (see description of function \texttt{modified.ttest}). These function types require a long computational time for
large images. An application with images of size $512\times 512$ will be discussed in Example 3.

The following R function was constructed to generate the covariance function $\bm{\Sigma}$. Prior to running the code we describe below, load the library
\textsf{SpatialPack} using the R command \texttt{library(SpatialPack)}.

\smallskip

{\small
\begin{lstlisting}[language=R]
Cov.nonsep <- function(h, u) {
  # Constructs the nonseparable covariance function
  psi <- function(r, a = 1, alpha = 1, b = 1)
     (a * r^alpha + 1)^b
  eta <- function(r, b = 1, g = 1, sigma = 1)
     (1 + (r/sigma^2)^g)^(-b/g)
  p <- 2
  r <- h^2 / psi(u^2)
  eta(r) / psi(u^2)^(p/2)
}
\end{lstlisting}
}

{\small
\begin{lstlisting}[language=R]
sim.images <- function(nsize, nobs = 10) {
  #  Simulates two correlated images
  coords <- expand.grid(1:nsize, 1:nsize)
  names(coords) <- c("xpos", "ypos")
  dmat <- as.matrix(dist(coords, diag = TRUE, upper = TRUE))
  S0 <- Cov.nonsep(dmat, u = 0)
  S1 <- Cov.nonsep(dmat, u = 1)
  Sigma <- rbind(cbind(S0, S1), cbind(S1, S0))
  p <- ncol(Sigma)

  # Generates 'nobs' random vectors from a multivariate
  # normal distribution
  Z <- matrix(rnorm(nobs * p), nrow = nobs, ncol = p)
  Z <- Z %*% chol(Sigma)

  # Computes the codispersion coefficient
  n <- nsize^2
  speed <- matrix(0, nrow = nobs, ncol = 5)
  for (i in 1:nobs) {
     x <- Z[i,1:n]
     y <- Z[i,(n+1):(2*n)]
     # Replace 'codisp' function by 'modified.ttest' if desired
     speed[i,] <- codisp(x, y, coords)$speed
  }

  # Creates the output
  speed <- speed[,-c(4,5)]
  colnames(speed) <- c("user","system","elapsed")
  list(speed = speed, times = apply(speed, 2, mean))
}
\end{lstlisting}
}

\subsection{Example 2: The Murray smelter site dataset}

\begin{figure}[!htp]
 \begin{center}
 \begin{minipage}[c]{6cm}
 \includegraphics[scale=.33]{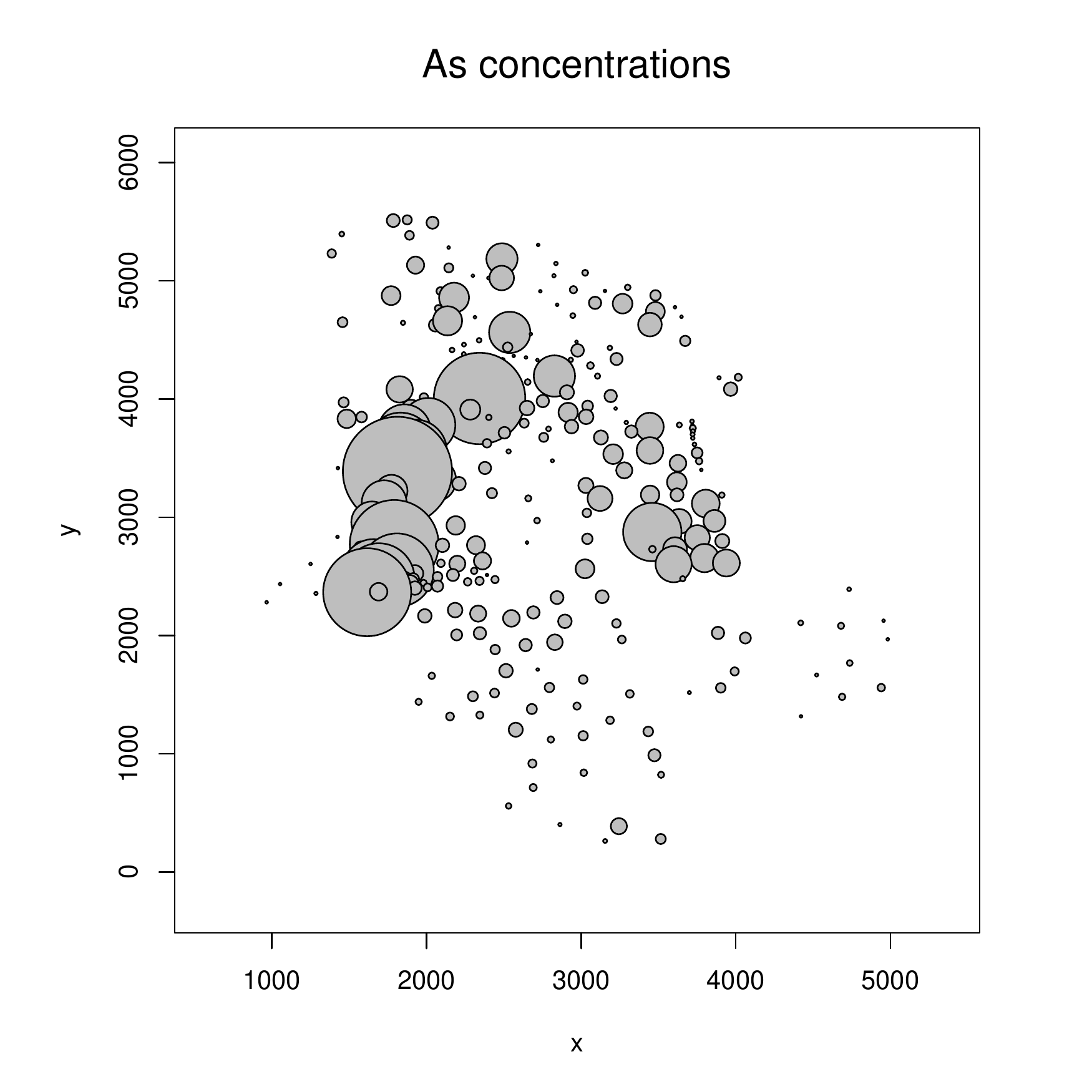}
 \end{minipage}
 \begin{minipage}[c]{6cm}
 \includegraphics[scale=.33]{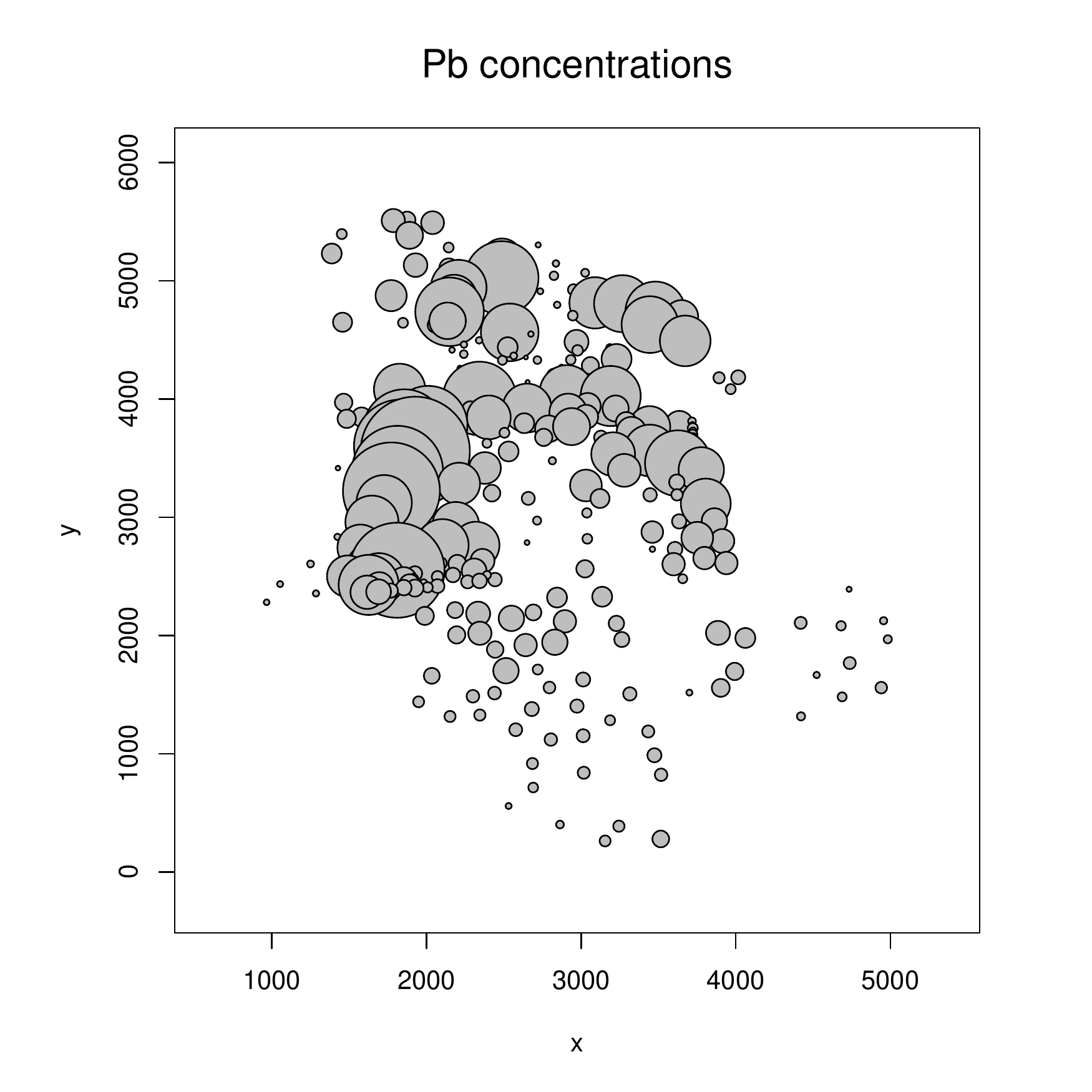}
 \end{minipage}
 \end{center}
 \caption{Bubble plots for As (left) and for  Pb (right).}\label{persp}
\end{figure}

Georeferenced data have been selected for illustrative purposes here. The dataset consists of soil samples collected in and around the vacant industrially
contaminated Murray smelter site (Utah, USA). This area was polluted by airborne emissions and the placement of waste slag from the smelting process. A total
of 253 locations were included in the study, and soil samples were taken from each location. Each georeferenced sampling quantity is a pool composite of four
closely adjacent soil samples for which the heavy metals arsenic (As) and lead (Pb) were measured. A complete description of the Murray smelter site dataset
can be found in \cite{Griffith:2002} and \cite{Griffith:2011}. The attributes As and Pb for each location are shown in Figure \ref{persp}.

The three methods described before were used to describe the spatial association between the variables As and Pb. The R output for the code
{\small
\begin{lstlisting}[language=R]
library(SpatialPack)
x <- murray$As
y <- murray$Pb
coords <- murray[,3:4]
murray.test <- modified.ttest(x, y, coords)
murray.test
\end{lstlisting}
}
\noindent%
provides $F=81.9490$, the degrees of freedom 1 and 154.0617 for the numerator and denominator, respectively, of the $F$ distribution, the $p\textrm{-value}=0$ and
the sample correlation coefficient $r=0.5893.$ Thus, the null hypothesis of no spatial association between the processes is rejected with a 5\% level of significance.
The code \texttt{summary(murray.ttest)} provides the upper boundaries for each of the thirteen (default) bins used in the computation of the modified $t$-test, and
for each class, the Moran coefficient is also given for both variables (As and Pb).

\begin{figure}
 \begin{center}
 \includegraphics[scale=0.35,angle=270]{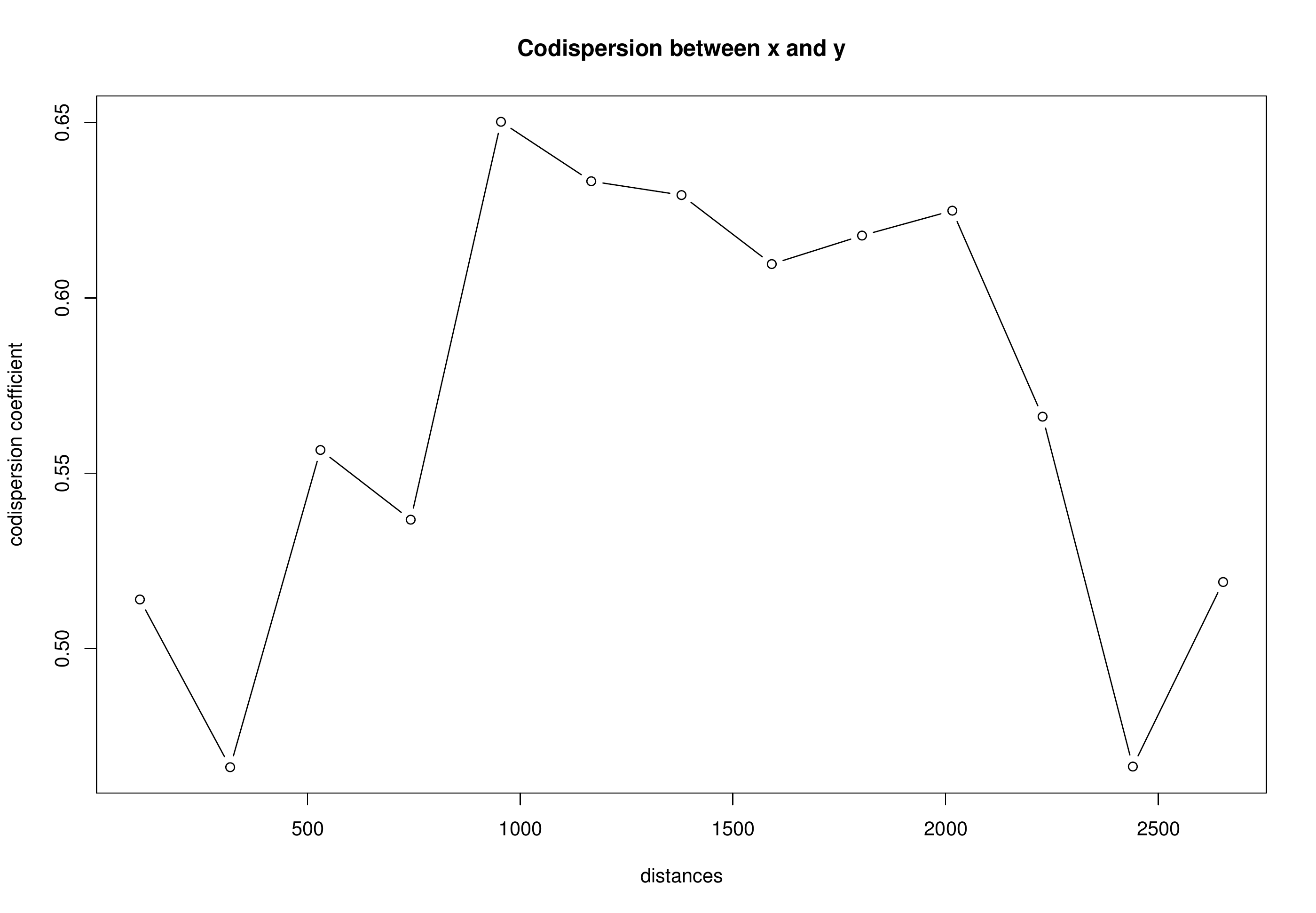}
 \end{center}
 \caption{Codispersion coefficient between As and Pb for the lag distance range from 0 to 2700 m.}\label{codis1}
\end{figure}

\noindent%
The following R code
{\small
\begin{lstlisting}[language=R]
murray.codisp <- codisp(x, y, coords)
murray.codisp
\end{lstlisting}
}
\noindent%
yields the codispersion for each of the bins. These codispersion values can be plotted as a function of the lag distance in the same way as the autocorrelation function
is displayed in time series. In Figure \ref{codis1}, $\widehat{\rho}_{XY}(h_k)$ versus $h_k$ has been plotted.
\noindent%
The maximum value of the codispersion coefficient (0.5602) is attained for a lag distance of 1000 m.

The R code
{\small
\begin{lstlisting}[language=R]
murray.cor <- cor.spatial(x, y, coords)
murray.cor
\end{lstlisting}
}
\noindent%
provides Tj{\o}stheim's coefficient and its variance. For the Murray dataset, the value of this coefficient is $-0.1519$, and its variance is $0.0035$. This nonparametric
correlation coefficient is far from the values of the other two methods, which provide evidence of positive spatial autocorrelation and a spatial codispersion in all
cases larger than $0.46$. These results agree with the findings reported by \cite{Rukhin:2008} in a series of Monte Carlo simulations in which the correlation coefficient,
Tj{\o}stheim's and the codispersion coefficient were compared in terms of bias and mean square error. For large values of the correlation between the processes, Tj{\o}stheim's
coefficient was the most biased coefficient.

\subsection{Example 3: Flammability of carbon nanotubes}

\begin{figure}[!htp]
 \begin{center}
 \begin{minipage}[c]{4.5cm}
 \includegraphics[scale=.25]{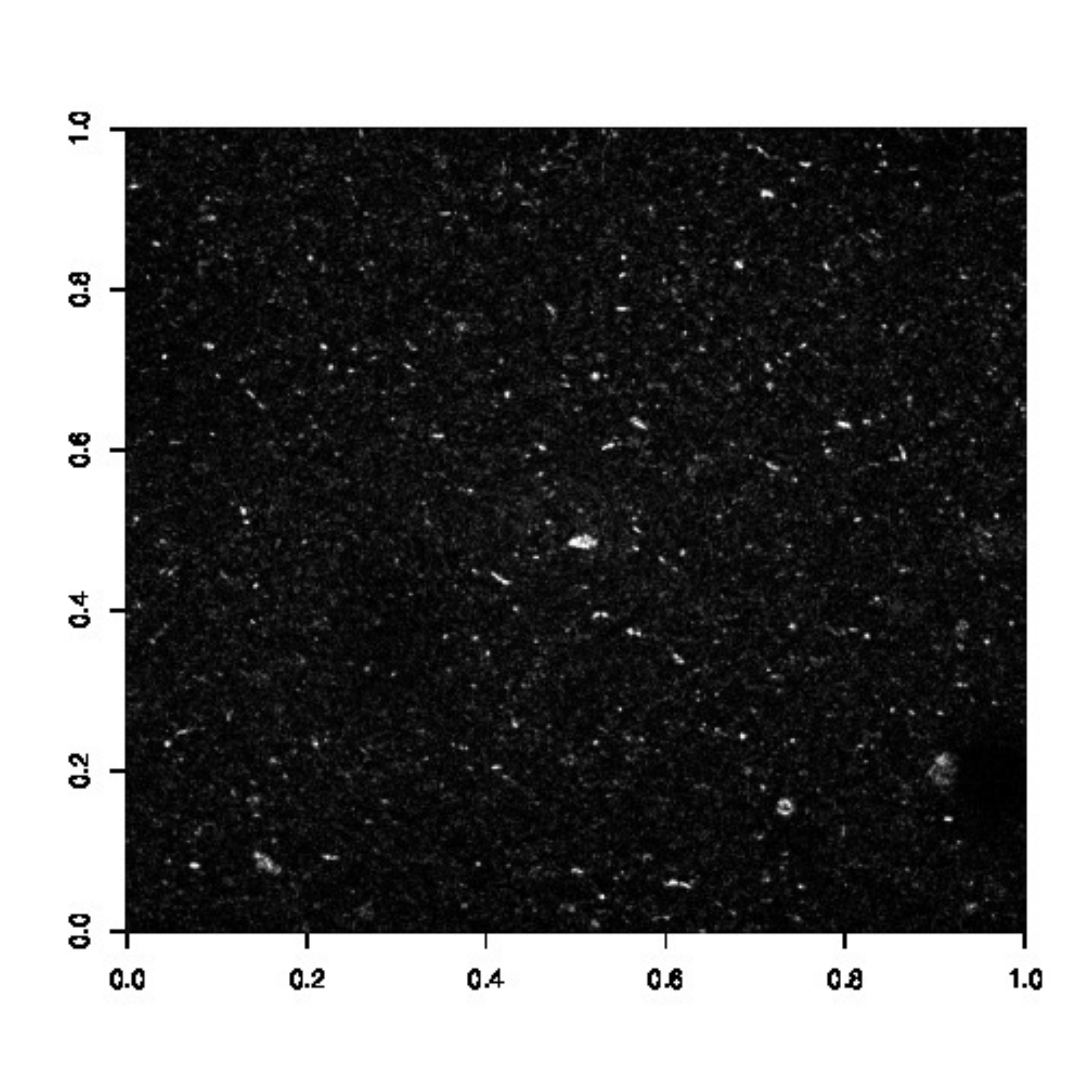}
 \centering
 (a)
 \end{minipage}
 \begin{minipage}[c]{4.5cm}
 \includegraphics[scale=.25]{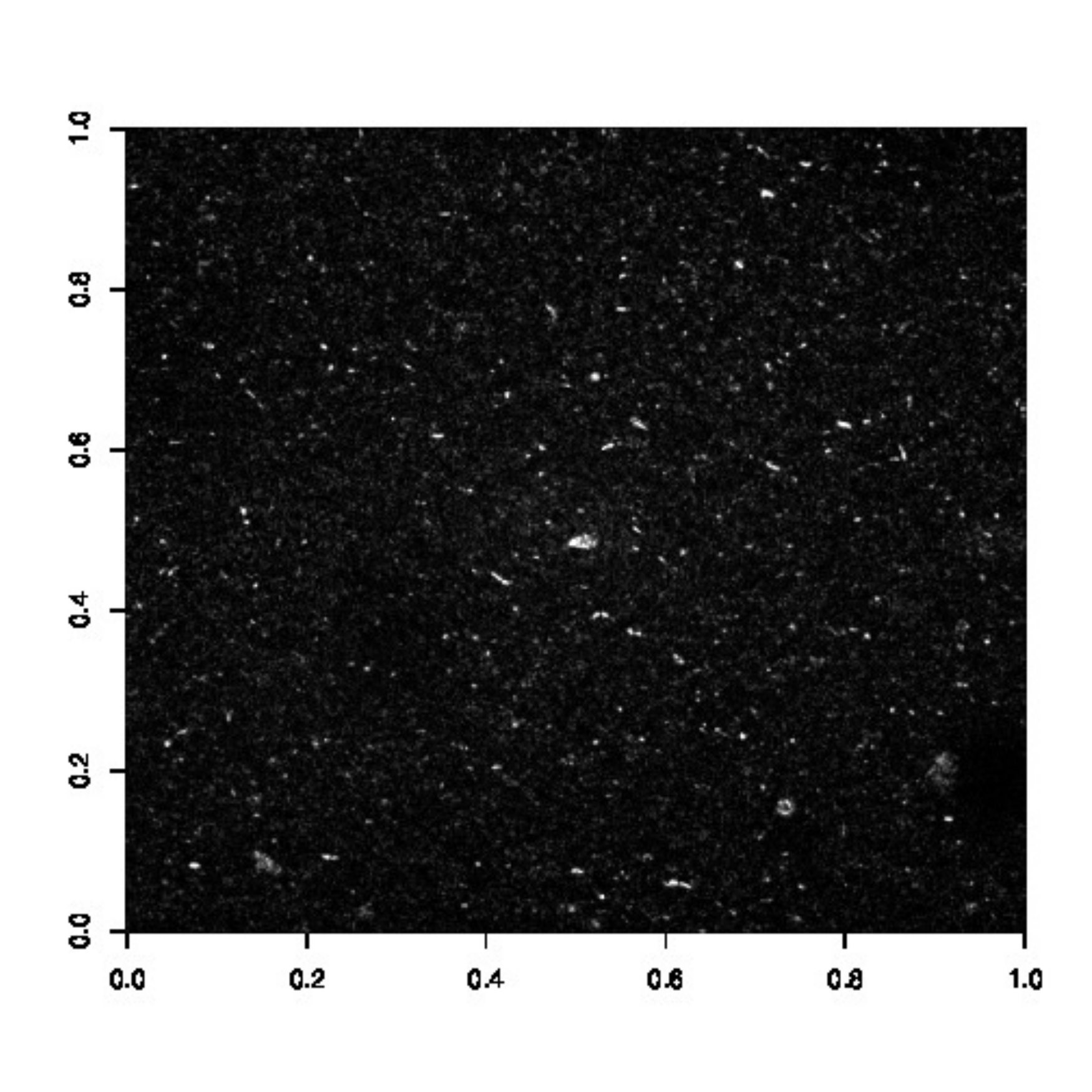}
 \centering
 (b)
 \end{minipage}
 \begin{minipage}[c]{4.5cm}
 \includegraphics[scale=.25]{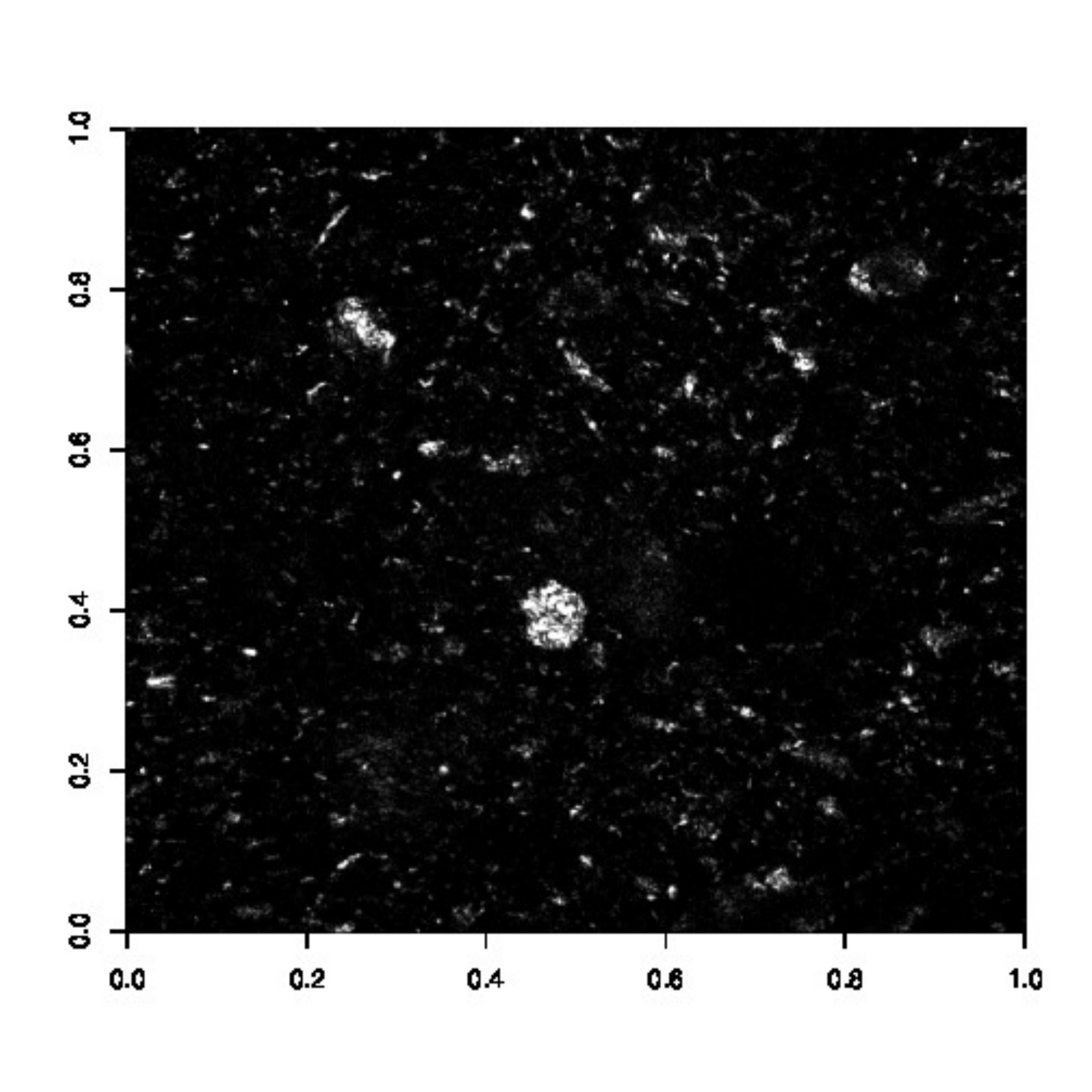}
 \centering
 (c)
 \end{minipage}
 \begin{minipage}[c]{4.5cm}
 \includegraphics[scale=.25]{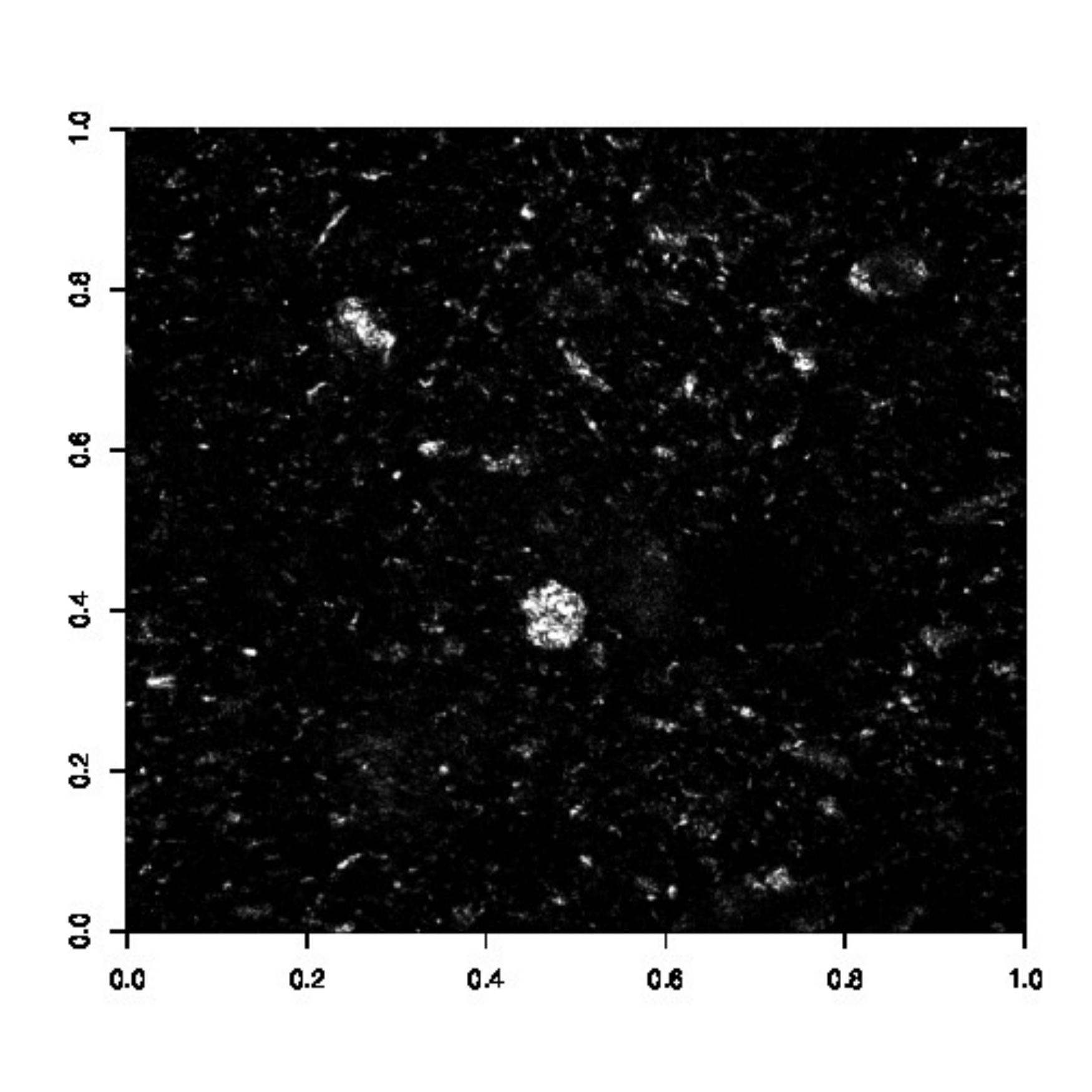}
 \centering
 (d)
 \end{minipage}
 \end{center}
 \vspace{-10pt}
 \caption{Dispersion of nanotubes.}\label{nanotubes}
\end{figure}

To illustrate a practical application of the codispersion coefficient with the data measured on a rectangular grid,
an example related to flammability of polymers is considered here. The flame retardant property of clay polymer
nanocomposites improve the physical and flammability properties of polymers \citep[see][]{Kashiwagi:2005}. The
distribution of this nanotube was examined by optical microscopy. This distribution is believed to depend mainly
on the distance from the top surface to the location of the polymer matrix (polymethyl methacrylate). The collected
dataset consists of four $512\times 512$ images plotted in Figure \ref{nanotubes}. Images (a) and (b) were
taken at the same distance from the main polymer matrix, while images (c) and (d) were both taken at the same distance
from the main polymer matrix but closer than images (a) and (b). \cite{Rukhin:2008} used the codispersion coefficient
as a metric of the closeness of the two sample polymers and fitted the spatial autoregressive processes to each image;
they computed the estimated codispersion coefficient between all pairs of images. As a result, the spatial association
between pairs (ac), (ad), (bc), and (bd) was almost null. However, the spatial association between the pairs (ab) and
(cd) was as high (greater than 0.8) as expected. These conclusions were obtained from the codispersion coefficient
$\rho_{XY}(\bm{h})$ that was evaluated in the directions $\bm{h}=(1,0) $, $\bm{h}=(0,1),$ and $\bm{h}=(1,1)$ \citep[see][]{Rukhin:2008}.

Here, we used the values provided by the function \texttt{codisp} to generate an omnidirectional plot $\rho(\|\bm{h}\|)$ against $\|\bm{h}\|$ in the same manner that the
omnidirectional variogram is plotted in spatial statistics. This process provides additional information about the codispersion range, and it is sometimes possible to
analyze the shape of the codispersion curve.
\begin{figure}[!htp]
 \begin{center}
 \includegraphics[scale=.45]{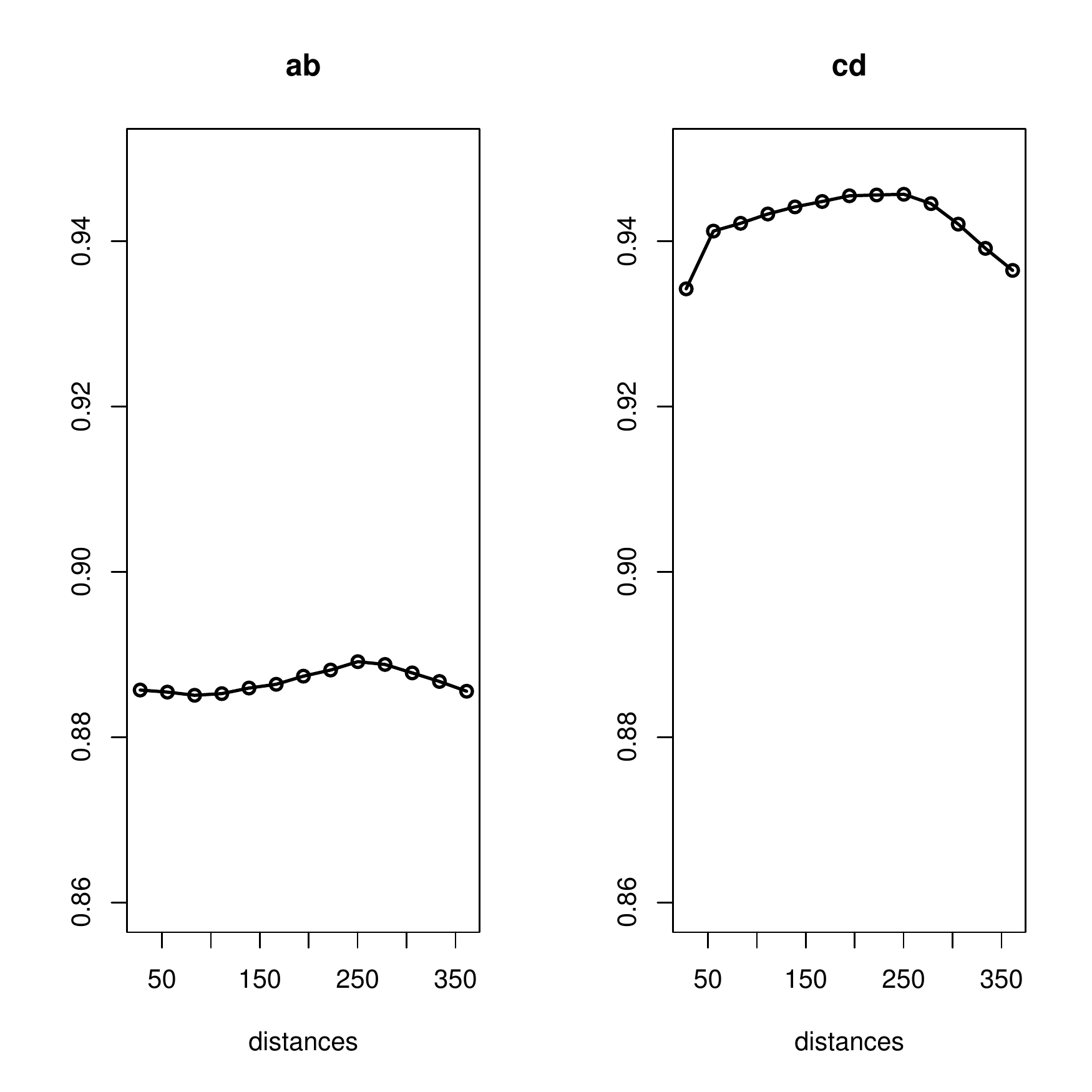}
 \end{center}
 \caption{Codispersion coefficient between the pairs of images (ab) (left) and (cd) (right).}\label{nanotubes1}
\end{figure}

\begin{figure}[!htp]
 \begin{center}
 \includegraphics[scale=.45]{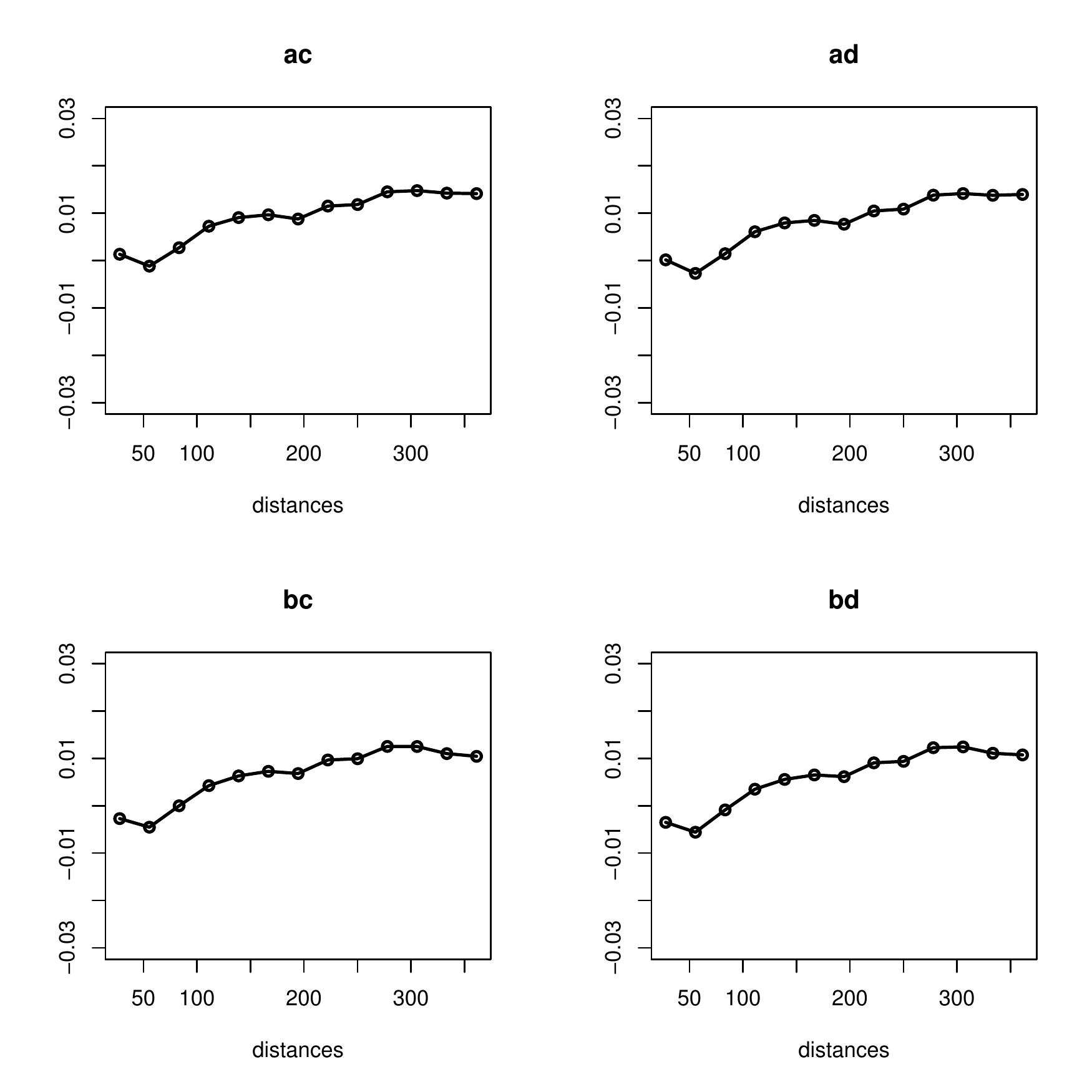}
 \end{center}
 \caption{Codispersion coefficient between the pairs of images(ac) (top left), (ad) (top right), (bc) (down left), and (bd)(down right)}\label{nanotubes2}
\end{figure}

Figure \ref{nanotubes1} shows the omnidirectional codispersion coefficient between the images pairs (ab) and (cd). In all cases, the values of the codispersion are greater
than 0.88 for images (a) and (b), and greater than 0.93 for images (c) and (d), which supports the findings of \cite{Rukhin:2008}. Additionally, the highest values of the
codispersion are attained for a distance of 250 units. Figure \ref{nanotubes2} shows the codispersion for the pairs of images (ac), (ad), (bc), and (bd). In all cases, the
values are close to zero.

The function takes the following computational times to compute the codispersion coefficients (considering the thirteen bins that are needed to yield the codispersion plot):
(ab) 5 hours, 45 minutes and 23 seconds; (ac) 5 hours 44 minutes and 8 seconds; (ad) 5 hours 43 minutes and 53 seconds; (bc) 5 hours 40 minutes and 42 seconds;
(bd) 5 hours 41 minutes and 2 seconds; (cd) 5 hours 41 minutes and 43 seconds. All computations were developed in a PC with a Core 2 quad q8400  2.66 Ghz processor,
and 8 Gb RAM DDR2 800 MHz.

\subsection{Example 4: Comovement between two time series}

The codispersion coefficient  as a measure of comovement between two time series was introduced by \cite{Vallejos:2008}. This coefficient is of interest in time series to
study how well two cronological sequences move together. When two spatial processes are defined on a set $D\subset\Rset^1$, the codispersion  is called  comovement
coefficient. It shares a number of the standard properties of the correlation coefficient, and is interpretable as the cosine of the angle between the vectors formed by the
first difference of the sample series. As in the case of the classic correlation, a comovement coefficient of $+1$ indicates that the sample function/processes being compare
are rescaled/retranslated versions of one another. Similarly, a profile matched with its reflection across the time axis gives a comovement of $-1$.

\textsf{SpatialPack} can be used  to quantify  the comovement between two time series. For illustrative purposes, a real data example is presented here. The data set
consist of two time series representing the monthly deaths from bronchitis, emphysema and asthma in the UK, 1974-1979. $X$ represents the males deaths while $Y$
represents the female deaths in the same period of time. The whole dataset is described in Table A.3 of \cite{Diggle:1990} and is available in R.

\begin{figure}[!htp]
 \begin{center}
 \includegraphics[scale=.45]{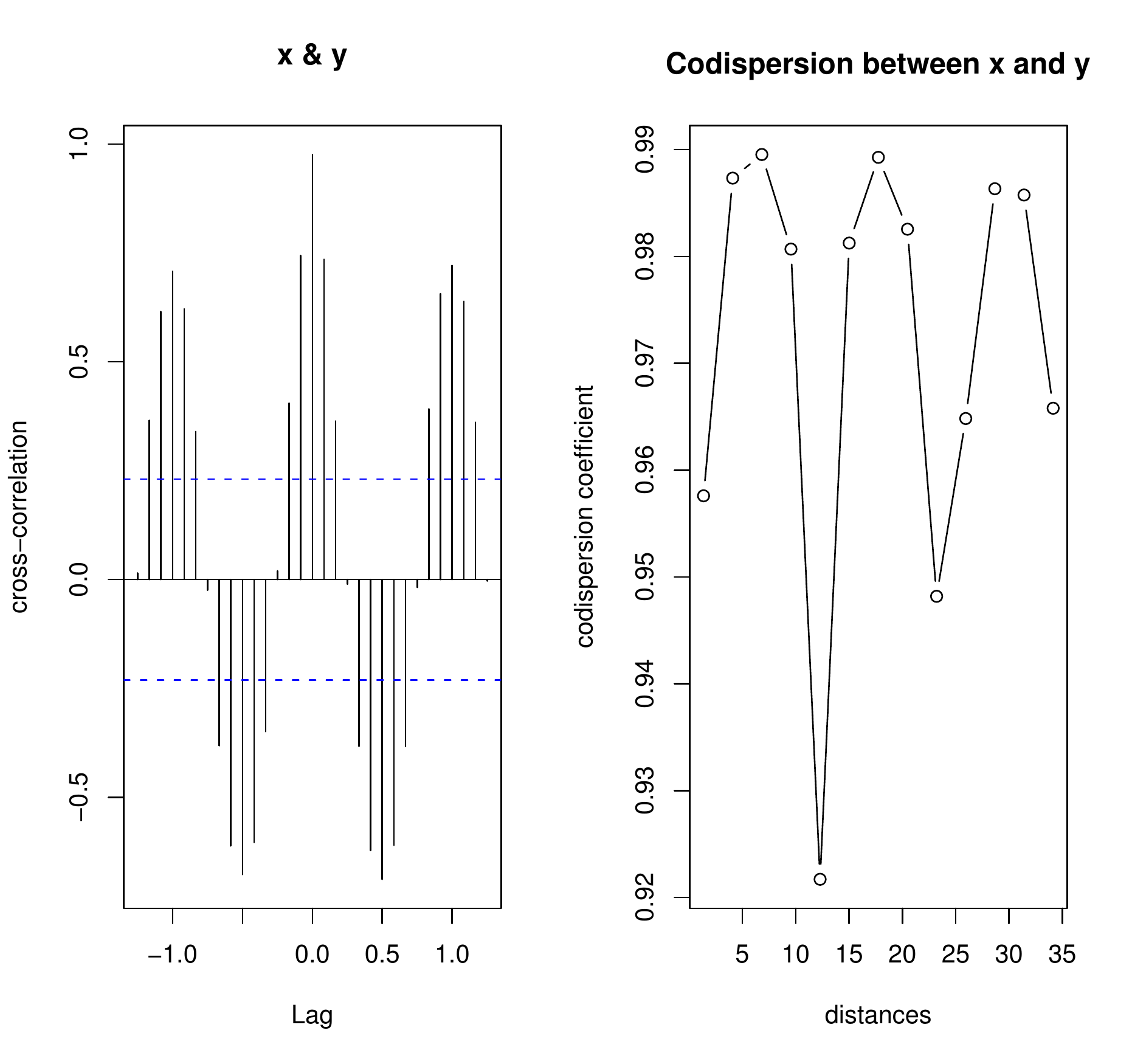}
 \end{center}
 \caption{Cross-correlation function (left) and comovement coefficient (right) between series $X$ and $Y$.}\label{t3}
\end{figure}

The R code
{\small
\begin{lstlisting}[language=R]
library(SpatialPack)
# UKLungDeaths from Package 'datasets'
x <- mdeaths
y <- fdeaths
coords <- cbind(1:72, rep(1,72))
z <- codisp(x, y, coords)
par(mfrow = c(1,2))
ccf(x, y, ylab = "cross-correlation", max.lag = 20)
plot(z)
\end{lstlisting}
}
\noindent%
produces Figure \ref{t3}. There is a seasonal pattern in the cross correlation function with significant components for certain values of the lag distance $h$.
The same behavior is observed for the comovement coefficient, where the maximum values are attained for $h=8$ and $h=19.$ For all values of $h$, $1\leq h \leq
35$, the comovement is greater than $0.9576$, showing a strong positive comovement between the males and females deaths in the period 1974-1979.

\section{Discussion}

In this paper we have presented \textsf{SpatialPack}, an R package, to assess the
association between two spatial processes. We provide three functions to explore the
spatial association between the two sequences. The package also provides interested
researchers and practitioners the facilities for working with the images and spatial
datasets that are defined on non-regular grids.

The current version of \textsf{SpatialPack} was tested for images sized $512 \times 512$
and for which the computational time is still reasonable (on average 5 hours and 40 minutes)
because the computationally intensive part was written in \texttt{C} and the storage of
the intermediate calculations was kept to a minimum.

\begin{figure}[htp]
 \begin{center}
 \includegraphics[scale=0.45]{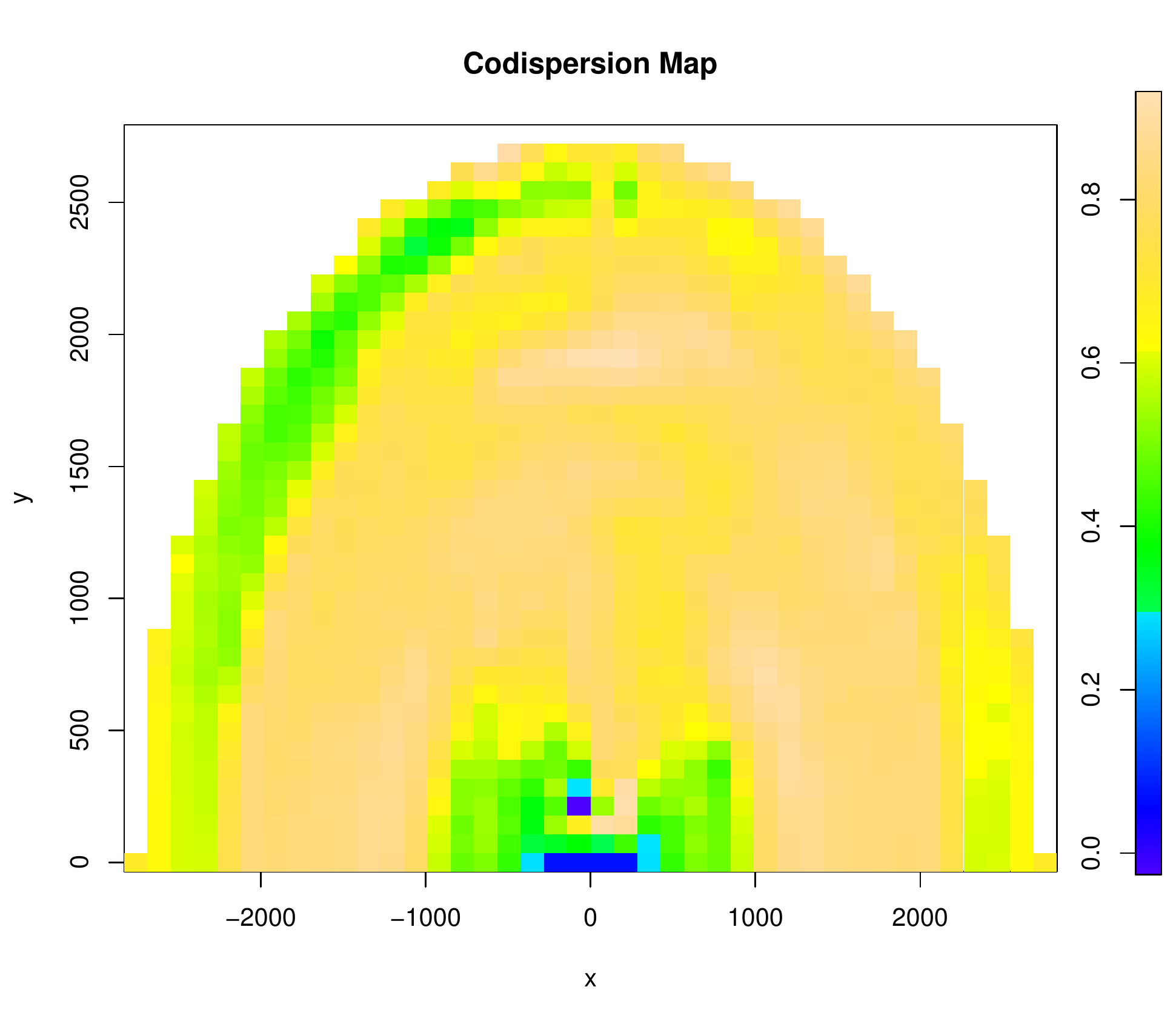}
 \end{center}
 \caption{Codispersion map for the murray smelter site dataset  of Example 2.}\label{codismap}
\end{figure}

Applications introduced illustrate how the spatial association between the two spatial
sequences can be analyzed in a number of different fields and contexts including time
series. Here, to emphasize that the codispersion coefficient provides information about
the direction for which the highest and lowest values are attained, we show a codispersion
map, a graphical display of the codispersion coefficient computed for all values of the
points belonging to a grid on the plane. The codispersion map is constructed computing
the method of moment estimator (\ref{eq:codisp}). As an example, for the Murray smelter
site dataset discussed in Example 2, a codispersion map is shown in Figure \ref{codismap}
for a circular grid that considers distances up to 3000 meters \citep[the code can be found in][]{Vallejos:2015b}.
\noindent%
The spatial correlation that is evident for an angle of 45$^\circ$ from Figure \ref{persp}
can be observed in Figure \ref{codismap} for values of the codispersion about $0.7$.

This package can be extended in several manners. For example, \cite{Cuevas:2013} studied a Nadaraya-Watson estimator of the codispersion coefficient. The R function,
to compute the nonparametric version of the codispersion coefficient, will be incorporated in future releases of \textsf{SpatialPack} including the codispersion map. On
the other hand, the generalization of the codispersion coefficient to address the problem of comparing more than two spatial processes has been recently considered.
In fact, \cite{Vallejos:2015a} uses the $m$-dimensional spatial vector $\bm{X}=(X_1(\bm{s}),X_2(\bm{s}),\dots,X_m(\bm{s}))^\top$, where $\bm{s}\in D$.
Let $X_i(\bm{s})$ be an intrinsically stationary process for all $l=1,2,\dots,m.$ For all $i,j=1,\dots,m$, and $\bm{h}\in D$ define
\begin{equation}\label{variog}
 \gamma_{X_i}(\bm{h}) = \E\left[\left(X_i(\bm{s}+\bm{h})-X_i(\bm{s})\right)^2\right],
\end{equation}
and
\begin{equation}\label{covariog}
 \gamma_{X_i X_j}(\bm{h}) = \E\left[\left(X_i(\bm{s}+\bm{h})-X_i(\bm{s})\right)\left(X_j(\bm{s}+\bm{h})-X_j(\bm{s})\right)\right].
\end{equation}
The variograms and cross-variograms given in equations (\ref{variog}) and (\ref{covariog}) were used to construct a codispersion matrix, called $\bm{\Psi}(\bm{h})$,
containing the codispersion coefficients between the components of $\bm{X}$. That is,
\[
 \bm{\Psi}(\bm{h})=\bm{\Gamma}_X^{-1/2}(\bm{h})\bm{\Gamma}_{XX}(\bm{h})\bm{\Gamma}_X^{-1/2}(\bm{h}),
\]
where $\bm{\Gamma}_{XX}(\bm{h})=(\gamma_{X_i X_j}(\bm{h}))$ and $\bm{\Gamma}_X(\bm{h})=\diag(\gamma_{X_i}(\bm{h}))$. The estimation of
$\bm{\Psi}(\bm{h})$ and the limiting distribution of the estimations were studied in the same way as in \cite{Rukhin:2008}. The computational aspects and applications
of the codispersion matrix will be included in future versions of \textsf{SpatialPack}.

Other routines related to spatial data analysis will be developed to further generalize the software. In addition, it will be of interest to develop specific methods to better
the connection between \textsf{SpatialPack} and other packages for spatial analysis, e.g. \textsf{geoR}.

\section*{Acknowledgments}

Ronny Vallejos was supported in part by FONDECYT  grant 1120048, Chile, and from AC3E,
grant FB-0008. Felipe Osorio was partially supported by FONDECYT grant 1140580. The
authors are indebted to Diego Mancilla for helpful discussions, and for the construction
of R routines for earlier versions of this article.


\end{document}